# Smart False Data Injection attacks against State Estimation in Power Grid

Muneer Mohammad, PhD., Member IEEE

*Abstract*— **In this paper a new class of cyber attacks against state estimation in the electric power grid is considered. This class of attacks is named false data injection attacks. We show that with the knowledge of the system configuration an attacker could successfully inject false data into certain state variable while bypassing existing techniques for bad data detection. In the preliminary section we consider the feasibility of such an attack and the necessary condition to successfully avoid detection. After that we show that with the knowledge of the system configuration, certain line flow measurements could be manipulated to lead to profitable misconduct. By controlling Regional Transmission Organizations (RTOs) view of system power flow and congestions, an attacker could manipulate the LMPs of targeted buses according to prior biddings.**

**Also, in this paper we show the implementation of the false data injection attacks. The numerical example considered was applied to a malicious data detection algorithm that was designed on a microcontroller. The results demonstrated the effectiveness of injecting false data measurements into the state estimation of electric power grids.**

***Keywords-; Power system state estimation, false data injection attacks, bad data detection, cyber attack, power system security***

## I. Introduction

The power grid is undergoing substantial changes as the idea of a smart grid is getting more popular. The implementation of the smart grid must be accompanied by numerous changes in the way the power grid is observed and analyzed. This is being made available by the technological advances in sensing, communication and actuation. While the majority of researchers have been involved in improving the power system operational efficiency and reliability, cyber attacks remain to be a vital threat to the power grid. Therefore, the potential risks of a cyber attack should be further studied before the implementation of a smart grid.

Power system researchers have realized the threat of bad data measurements and developed certain techniques for processing them e.g., [1]-[3]. The techniques developed involve two steps, first detect if there are bad measurements and then identify and remove them. This technique as well as other techniques is aimed at filtering arbitrary bad measurements. If any smart attack is considered as random bad data, then the developed methods should be able to successfully filter the false measurements.

However, this paper shows that the developed bad data detection techniques could be bypassed if a certain smart attack is injected into the system. Nevertheless, the construction of certain attacks requires knowledge of the power grid configuration. The reason that that this attack could be made is the assumption that: "when bad measurements are present, the squares of differences between the observed measurements and their corresponding estimates become significant [1]." Therefore, the attacker can bypass bad measurement detection by violating this assumption. This can be carried out with the proper knowledge of the power grid configuration.

The operation of the power grid became significantly different after the industry deregulation. The operation of the physical power systems are organized into Regional Transmission Organizations (RTOs). As a result, a potential cyber attack on the state estimation of these RTOs has recently become a major concern. The Energy Management System (EMS) is responsible to relay a robust estimation of the entire system. Therefore, this management system will be a major target for attacker aiming to manipulate the system. In [4] the scheme of false data injection attacks against a power grid's state estimation was first introduced. The paper showed that by acquiring certain information about the network topology the bad data detection techniques employed by SCADA systems could be bypassed. In [5] two possible indices are proposed in order to quantify required efforts to implement such a class of attack. In [6] there were attempts to develop computationally efficient heuristics to detect such false data attacks against state estimators. In [7] the method of employing such attacks to lead to profitable financial misconduct was presented. Such profit could be acquired by virtual bidding the ex-post LMP while simultaneously injecting false data into the state estimation.

In this paper we present false data injection attacks from the attacker's perspective. We first show that the bypassing of bad data detection could be achieved. We then apply the presented injection attacks onto a 5-bus system. Three scenarios are studied on the same system. In the first scenario actual measurements are taken directly to the system without bad data injection. Also, two methods were used to detect bad data: Normalized Residuals and Chi-Square methods. In the second scenario, bad data is injected into the system using the attack scheme presented. The attack scheme aims at injecting undetectable random data into the system. It is then shown that both methods of detection fail to identify the presence of the injected bad measurements. In the third scenario random bad data is added to the system to demonstrate the effectiveness of bad data detection techniques of malicious bad data.

The rest of this paper is organized as follows. Section 2 gives background and shows some related work. Section 3 presents the basic principle of false data injection attacks. Section 4 demonstrates the success of these attacks by simulation. Section 5 concludes this paper points out future research directions.

## II. PRELIMIARIES

### A. Power System (Power Grid))

The power transmission system consists of electric generators, transmission lines, and transformers that form a power grid. It connects a variety of electric generator together with a host of users across a large geographical area. Redundant paths and line are provided so that power can be routed from an y power plant to any customer, through a variety of routes, based on the economics of transmission paths and cost of power. A control center is usually used to monitor and control the power system and devices in a geographical area.

### B. State Estimation Background

It is essential to monitor and control the power flows and voltages in a power system in order to maintain system reliability and stability. Power meters scattered around the power grid provide control centers with the necessary measurements to monitor the system. These measurements are used in the state estimation algorithm of the EMS to estimate the state of the system variables. State variables include bus voltages and magnitudes.

The DC power flow model of a state estimation problem is most commonly used. This is a linearization of the AC power flow problem. The goal is to estimate the power system variables $x \in R^n$ based on measurements $z \in R^m$. Therefore, x and z are related according to

$$z = Hx + e \quad (1)$$

where $H = (h_{i,j})_{m \times n}$. The statistical estimation criteria most commonly used in the weighted least-square criterion. When meter error is assumed to be normally distributed with zero mean, this criteria leads to an identical estimator with the following matrix solution

$$\hat{x} = (H^T R^{-1} H)^{-1} H^T R^{-1} z \quad (2)$$

where R is the inverse of W; a diagonal matrix whose elements are reciprocals of the variances of meter error.

$$R = W^{-1} \quad (3)$$

$$W = \begin{bmatrix} \sigma_1^{-2} & & & & \\ & \sigma_2^{-2} & & & \\ & & . & & \\ & & & . & \\ & & & & \sigma_m^{-2} \end{bmatrix} \quad (4)$$

where $\sigma_i^2$ is the variance of the i-th meter $(1 \le i \le m)$.

### C. Bad Data Measurment Detection

Bad measurements could be present due to several reasons such as meter failure, miscommunication and malicious attacks. There are certain techniques that have been developed to detect these bad data and remove them to protect the robustness of the state estimator [8]. In a regular measurement, a meter provides an estimate of the state variable close to the actual value, while abnormal measurements are skewed from their true values. Researchers have proposed to calculate the measurement residual $r = z - H\hat{x}$ (i.e. the difference between observed measurements and estimated measurements) and then use its L$_2$-norm $\| z - H\hat{x} \|$ to detect the presence of bad data. Essentially, $\| z - H\hat{x} \|$ is compared to a threshold $\tau$, if $\| z - H\hat{x} \| > \tau$, then there is bad data.

The Normalized Residual method takes into account the maximum normalized residual and compares that with a threshold determined by a unit Gaussian function to determine the presence of bad data. Also it can be shown mathematically that $\| z - H\hat{x} \|^2$ follows a $\chi^2(v)$-distribution (chi-square), where $v = m - n$ is the degree of freedom. This method is called the Chi-Square method of detecting bad data thus the threshold $\tau$ is determined using a Chi-Square function.

A level of certainty is associated with each choice of $\tau$. This level of certainty is chosen by the system operator to represent the level of trust we associate to our bad data detector. In this paper a certainty level of 99% was chosen.

## III. FALSE DATA INJECTION ATTACK MODEL

In this section we will show how smart false data injection could be carried out to bypass the bad data detection methods introduced earlier. Let us consider a system with *n* state variables $x_1, ..., x_n$ and *m* meters with measurements $z_1, ..., z_m$. The **H** matrix described in Section 2 is then an m x n matrix that describes the relationship between the measurements and the state variables. This **H** matrix is a constant matrix determined by the placement of the meters and the line impedances. In [8] it is described how control centers construct **H**. In our application we assume that the attacker has access to this **H** matrix as well as access to the meters to be compromised.

In the first part we discuss the smart random false data injection, meaning the attacker aims at injecting false data into random meters. Also, targeted false data is presented in which the attacker aims at altering a specific measurement for certain benefit.

In the following sub-sections we explain the mathematical principal of false data injection and how it is able to bypass detection. Then we apply the attack of each scenario.

## A. Basic Mathematical Principle

As defined earlier $z = (z_1,...,z_m)^T$ is the original vector of measurements without any alteration. Then we define a vector $z_a$ that represents the vector of observed measurements that contains malicious data. Therefore, $z_a$ can be represented as $z_a = z + a$, where $a = (a_1,...,a_m)^T$ is the attack vector that contains the malicious data added to the original measurements.

Also, let $\hat{x}_{bad}$ and $\hat{x}$ represent the estimates of $x$ using malicious data $z_a$ and the original measurements $z$, respectively. Then we can represent $\hat{x}_{bad} = \hat{x} + c$, where $c$ is a non-zero vector of length $n$. This vector $c$ then represents the error injected by the attacker.

Therefore, the attackers objective would be to inject an attack vector using **Hc** (i.e. a=Hc). Since bad data detection methods compute the $L_2$-norm of the measurement residual to check for the existence of bad data. The $L_2$-norm of the malicious data is the same as that of the original measurements if an attack vector is chosen in that way. In other words, choosing an attack vector that is a linear combination of the columns of H would bypass the detection technique that utilizes the norm of the residuals. Theorem 1 below shows the proof of the above claim.

*THEOREM* 1. *If the original measurements z can bypass bad data detection then the malicious measurements $z_a = z + a$ can pass the detection test if **a** is a linear combination of the column vectors of **H** [4].*

*PROOF.* Since z can pass the detection, we have $\| z - H\hat{x} \| < \tau$, where $\tau$ is the detection threshold assigned. Also, if we have **a** = **Hc** then the norm of the measurement residual is,

$$\| z_a - H\hat{x}_{bad} \| = \| z + a - H(\hat{x} + c) \|$$
$$= \| z - H\hat{x} + (a - Hc) \|$$
$$= \| z - H\hat{x} \| \leq \tau \quad (5)$$

Thus, the norm of the measurement residual of $z_a$ is also less than the threshold $\tau$. This implies that $z_a$ can also bypass the bad data detection process. The attacker can choose a random vector **c** to inject random false data into the system; alternatively a carefully calculated vector **c** could change specific measurements to a desired value.

## B. Random False Data Injection

We assume the attacker has access to k meters in the power grid. Also assume $I_m = \{i_1,...,i_k\}$ is the set of indices of those meters. Then the attacker needs to create an attack vector $a = (a_1,...,a_m)^T$ such that $a_i = 0$ for $i \notin I_m$ and **a** is a linear combination of the column vectors of **H**. There are several methods to calculate the attack vector **a**. [8] presents a method that rewrites the relation **a** = **Hc** as **Ba=0**. Where, $B = P - I$ and P is defined as $P = H(H^T H)H^T$. In [4] it is also shown that if $k \geq m - n + 1$ then there exists an attack vector **a** in which $a_i = 0$ for $i \notin I_m$.

**Construction of Attack Vectors:** When the condition $k \geq m - n + 1$ is satisfied the attacker can directly perform column transformations on H, such that the resulting matrix is a linear combination of the columns. However, the elements that are not controlled by the attacker have to be eliminated. Then each such constructed vector could be used as an attack vector.

## IV. NUMERICAL EXAMPLE

In this section, we confirm that false data injection attacks could be injected into state estimators and remain undetected by bad data detection techniques. Also, we show that false data could be injected to target certain measurements to affect the Locational Marginal Price (LMP) at certain nodes. A 5-bus system with 6 branch measurements was used as our test system (figure 1). In the first part of the simulation the goal is to show the feasibility of creating an attack vector that could bypass bad data detection.

In our test case, a DC power flow model was assumed; all nodes have a voltage level of 1p.u. Also, the measurements are real power branch flow measurements. The **H** matrix was derived according to the assumed line impedances. State estimation and bad data detection were simulated using MATALB R2008a, on the other hand LMP simulations were carried out in PowerWorld.

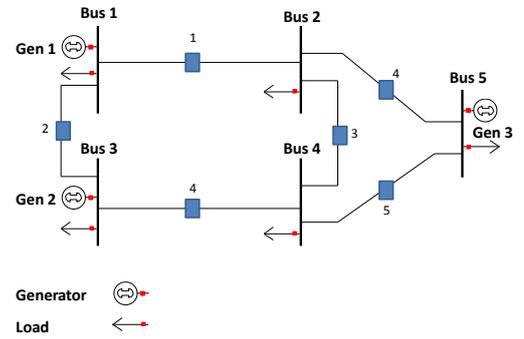

Figure 1: 5-Bus test system

**Table I. Line impedance and Original Measurements**

| Measurement (m) | Branch | Line Impedance (p.u.) | Measurements Values (p.u.) |
|---|---|---|---|
| 1 | 1-2 | 0.03 | 0.91 |
| 2 | 1-3 | 0.05 | -0.16 |
| 3 | 2-4 | 0.05 | 0.19 |
| 4 | 2-5 | 0.08 | 0.21 |
| 5 | 3-4 | 0.05 | 0.89 |
| 6 | 4-5 | 0.08 | 0.09 |

## A. False Data Injection Results

The case system used is shown in figure 1 and the line impedances are given in Table I. Three cases are applied to the 5 bus system. Case 1 is where original measurements are applied to the system without any interference. In this case a normal optimal power flow is solved with LMP considerations. Therefore, when bad data detection is applied the estimator shows that the estimated measurements are clean of malicious data. In case 2, a random vector is applied to the original measurement vector. The state estimator functions properly in estimation; as well as, the bad data detection which reveals the false data and reveals which branch measurement should be dropped from the estimator. Two bad data detection techniques were used: Normalized residuals and Chi-Square method (discussed earlier). In case 3, smart false data is injected into the measurement vector. This attack vector is randomly chosen with the only condition is to satisfy **a = Hc**. The attack vector successfully alters the estimated measurements and more importantly bypasses both bad data detection techniques applied.

*B. False Injection for Profit*

In the second part of the example, the ability of the attacker to manipulate the market price based on false data injection was tested. The 5-bus system was modeled in PowerWorld with the appropriate line properties to maintain a constant **H** matrix. The loads were also designed to ensure the real power line flows most accurately resembled the original measurements taken. The generator properties and load magnitudes are given in Tables II &III. In the initial run of market clearing prices with OPF constraints the results were that Gen 1. would generate at maximum capacity while Gen 2. would pick up the remaining load and Gen 3. was kept on reserve.

The false data injection was simulated in Matlab to obtain the new (false) measurement estimates. The **c** vector was calculated to simulate congestion on line 3-4. This would then limit the power production by the generator on bus 3. Then to model this change in measurements the load was manipulated in PowerWorld to resemble the change. After that, the market clearing prices were calculated and the result was that Gen 3. was forced to operate, thus increasing the LMP at buses 2,4 and 5. The change in LMP is shown in Table IV. Then the attacker would buy power at bus 1 before injection and sell power at bus 4 after false data injection for maximum profit. This manipulation of the market clearing prices would remain to be undetectable because of the inability of the bad data detection to sense the erroneous measurements.

**Table II. Generator Properties and Load Magnitudes**

| Gen number | Price ($/MWh) | Maximum Capacity (MW) |
|---|---|---|
| 1 | 10 | 250 |
| 2 | 15 | 300 |
| 3 | 30 | 500 |

**Table III. Load Magnitudes**

| Load on Bus number | Load Magnitude (MW) |
|---|---|
| 1 | 100 |
| 2 | 100 |
| 3 | 200 |
| 4 | 40 |
| 5 | 60 |

**Table IV. LMP Before and After False Data Injection**

| Bus number | LMP Before False Injection ($/MWh) | LMP After False Injection ($/MWh) |
|---|---|---|
| 1 | 15 | 22.57 |
| 2 | 15 | 27.12 |
| 3 | 15 | 15 |
| 4 | 15 | 32.88 |
| 5 | 15 | 30 |

V. HARDWARE IMPLEMENTATION

In this section, the hardware implementation of our system design is presented and discussed. The realization of such a system was designed on Proteus. The main component that was used in our design was a PIC microcontroller. The source code is written on a PC host using a C compiler, assembled and downloaded to the chip as shown in Fig. 2.

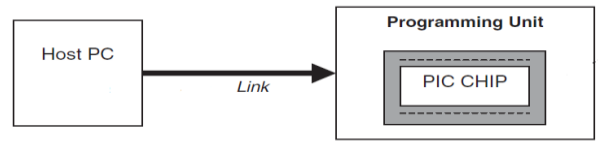

Fig. 2. Programming Operations of the PIC

A serial link is the connection unit that serves as a delivery communication between the host PC and the programming unit. The PC runs a C-compiler, so that once a program has been written and assembled, it is downloaded by placing the chip in a PICSTART Plus connected to a PC [9]. Fig. 3 shows the proposed design for the PIC microcontroller that has been used in this project.

The objective of the PIC in this project is to identify the presence of the injected bad measurements based on the simulation has been done by the Matlab

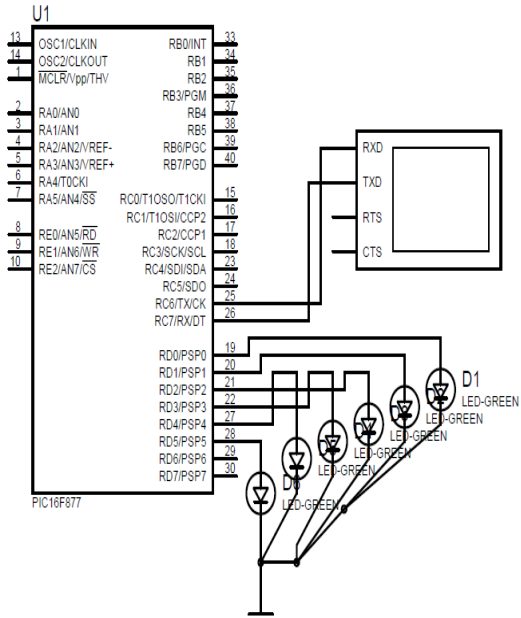

Fig. 3   Proteus Simulation of the Circuit Design

In our design, a 20 MHz crystal circuit is used to give a 200ns instruction execution time and to create a precise frequency when subject to electrical stimulation, which is more convenient for analyzing program timing [10]. Moreover, the crystal oscillator needs to be physically near the MCU to prevent the crystal from oscillating, or affecting the resonant frequency. In addition, a 10pF capacitor is used to avoid the clock signal causing a malfunction, and two 15Pf are used to stabilize the frequency.

As shown in Fig. 3, the Proteus simulation software was used to write in C-compiler in order to provide virtual instruments which was used to build up the circuit, just as in a real application

## VI. CONCLUSION

In this paper we present a potential cyber attack on the state estimation system used in Regional Transmission Organizations (RTOs). The SCADA system uses DC power flow model to estimate the state of the system using real measurements. We show how an attacker could inject false data into the state estimation system while bypassing bad data detection techniques. Also, we show how an attacker could target specific branches to change the LMP on certain nodes for financial gains.

Some future directions of this work would be to design a more robust state estimator that could detect such false data injection attacks. Some efforts have been done to design indices to show how we can limit such attacks [8]. Also, a counter-attack method that employs a linear combination of **H** could also be designed to stop the effect of such attacks. Redundancy methods would also offer a solution to such a problem of malicious injected data.

REFERENCES

[1]  Jeu-Min Lin; Heng-Yau Pan; , "A Static State Estimation Approach Including Bad Data Detection and Identification in Power Systems," *Power Engineering Society General Meeting, 2007. IEEE* , vol., no., pp.1-7, 24-28 June 2007.

[2]   Amr A. Munshi, Yasser Abdel-Rady I. Mohamed, "Data Lake Lambda Architecture for Smart Grids Big Data Analytics", Access IEEE, vol. 6, pp. 40463-40471, 2018.

[3]   R. M. Lee, M. J. Assante, T. Conway, *Analysis of the cyber attack on the ukrainian power grid*, 2016..

[4]  Y. Liu, M. K. Reiter, and P. Ning, "*False data injection attacks against state estimation in electric power grids*," in Proc. of the 16th ACM *Conference on Computer and Communications Securty 2009*.

[5]   H. Sandberg, A. Teixeira, and K. H. Johansson, "On security indices for state estimators in power networks," in *Preprints of the First Workshop on Secure Control Systems, CPSWEEK 2010*, 2010.

[6]  Kosut, O.; Liyan Jia; Thomas, R.J.; Lang Tong; , "Limiting false data attacks on power system state estimation," *Information Sciences and* Systems (CISS), 2010 44th Annual Conference on , vol., no., pp.1-6, 17-19 March 2010

[7]  G. Chaojun, P. Jirutitijaroen, M. Motani, "Detecting false data injection attacks in ac state estimation", IEEE Transactions on Smart Grid, vol. 6, no. 5, pp. 2476-2483, Sept 2015

[8]  A. Monticelli.  State Estimation in Electric Power Systems, A Generalized Approach. Kluwer Academic Publishers, 1999.

[9]  M. Battes, Interfacing PIC Microcontroler, Jan. 2007, vol. 35, no. 1.

[10]  J. Iovine, PIC microcontroller project book. McGraw-Hill, Inc. New York, NY, USA, 2004.